# Physics and application of photon number resolving detectors based on superconducting parallel nanowires

F Marsili[1,2*], D Bitauld[1], A Gaggero[3], S Jahanmirinejad[1], R Leoni[3], F Mattioli[3] and A Fiore[1]

[1]COBRA Research Institute, Eindhoven University of Technology, P.O. Box 513, NL-5600MB Eindhoven, The Netherlands
[2]Ecole Polytechnique Fédérale de Lausanne (EPFL), Institute of Photonics and Quantum Electronics (IPEQ), Station 3, CH-1015 Lausanne, Switzerland;
[3]Istituto di Fotonica e Nanotecnologie (IFN), CNR, via Cineto Romano 42, 00156 Roma, Italy;

*Corresponding author. Email: **francesco.marsili@epfl.ch**

**Abstract.** The Parallel Nanowire Detector (PND) is a photon number resolving (PNR) detector which uses spatial multiplexing on a subwavelength scale to provide a single electrical output proportional to the photon number. The basic structure of the PND is the parallel connection of several NbN superconducting nanowires (≈100 nm-wide, few nm-thick), folded in a meander pattern. PNDs were fabricated on 3-4 nm thick NbN films grown on MgO ($T_S$=400°C) substrates by reactive magnetron sputtering in an Ar/$N_2$ gas mixture. The device performance was characterized in terms of speed and sensitivity. PNDs showed a counting rate of 80 MHz and a pulse duration as low as 660ps full width at half maximum (FWHM). Building the histograms of the photoresponse peak, no multiplication noise buildup is observable. Electrical and optical equivalent models of the device were developed in order to study its working principle, define design guidelines, and develop an algorithm to estimate the photon number statistics of an unknown light. In particular, the modeling provides novel insight of the physical limit to the detection efficiency and to the reset time of these detectors. The PND significantly outperforms existing PNR detectors in terms of simplicity, sensitivity, speed, and multiplication noise.

Equation Chapter (Next) Section 1

## 1. Introduction

Photon number resolving (PNR) detectors are required in the fields of quantum communication, quantum information processing and of quantum optics for two class of applications. In one case PNR detectors are needed to reconstruct the incoming photon number statistics by ensemble measurements. This is the case of the characterization of nonclassical light sources such as single photon [1] or *n*-photon [2] state generators or of the detection of pulse splitting attacks in quantum cryptography, where an eavesdropper alters the photon statistics of the pulses [3]. In the second case the PNR detectors are needed to perform a single-shot measurement of the photon number. Applications of this kind are the linear-optics quantum computing [4], long distance quantum communication (which requires quantum repeaters [5]) and conditional-state preparation [6].

Among the approaches proposed so far to PNR detection, detectors based on charge-integration or field-effect transistors [7-9] are affected by long integration times, leading to bandwidths <1 MHz. Transition edge sensors (TES [10]) show extremely high (95%) detection efficiencies but they operate at 100 mK and show long response times (several hundreds of nanoseconds in the best case). Approaches based on photomultipliers (PMTs) [11] and avalanche diodes (APDs), such as the visible light photon counter (VLPC) [2, 12], 2D arrays of APDs [13, 14] and time-multiplexed detectors [15, 16] are not sensitive or are plagued by high dark count rate and long dead times in the telecommunication spectral windows. Arrays of SPDs additionally involve complex read-out schemes [14] or separate contacts, amplification and discrimination [17]. We recently demonstrated an alternative approach [18, 19], the Parallel Nanowire Detector (PND), which uses spatial multiplexing on a subwavelength scale to provide a single electrical output proportional to the photon number. The device presented significantly outperforms existing PNR detectors in terms of simplicity, sensitivity, speed, and multiplication noise. Here we present the working principle of the device (section 2), a review of fabrication and of experimental results (section 3 to 5), an extensive analysis of the device operation and corresponding design guidelines (section 6) and the first application of a PND to reconstruct an unknown incoming photon number statistics (section 7).

## 2. Photon Number Resolution principle

The structure of PNDs is the parallel connection of N superconducting nanowires (N-PND), each of which can be connected in series to a resistor $R_0$ (N-PND-R, figure 1b). The detecting element is a 4-6 nm thick, 100 nm wide NbN wire folded in a meander pattern. Each section acts as a nanowire superconducting single photon detector (SSPD) [20]. In SPPDs, if a superconducting nanowire is biased close to its critical current, the absorption of a photon causes the formation of a normal barrier across its cross section, so almost all the bias current is pushed to the external circuit. In PNDs, the currents from different sections can sum up on the external load, producing an output voltage pulse proportional to the number of photons absorbed.

The time evolution of the device after photon absorption can be simulated using the equivalent circuit of Figure 1a. Each section is modeled as the series connection of a switch which opens on the hotspot resistance $R_{hs}$ for a time $t_{hs}$, simulating the absorption of a photon, of an inductance $L_{kin}$, accounting for kinetic inductance [21] and of a resistor $R_0$. The device is connected through a bias T to the bias voltage source $V_B$ and to the input resistance of the preamplifier $R_{out}$. The n firing sections, in pink, all carry the same current $I_f$ and the N-n still superconducting sections (unfiring), in green, all carry the same current $I_u$. $I_{out}$ is the current flowing through $R_{out}$.

Let $I_B$ be the bias current flowing through each section when the device is in the steady state. If a photon reaches the i$^{th}$ nanowire, it will cause the superconducting-normal transition with a probability $\eta_i=\eta(I_B/I_C^{(i)})$, where η is the current-dependent detection efficiency and $I_C^{(i)}$ is the critical current of the nanowire [20] (the nanowires have different critical currents, being differently constricted [22]). Because of the sudden increase in the resistance of firing nanowire, its current ($I_f$) is then redistributed between the other N-1 unfiring branches and $R_{out}$. This argument yields that if n sections fire simultaneously (in a time interval much shorter than the current relaxation time), part of their currents sum up on the external load.

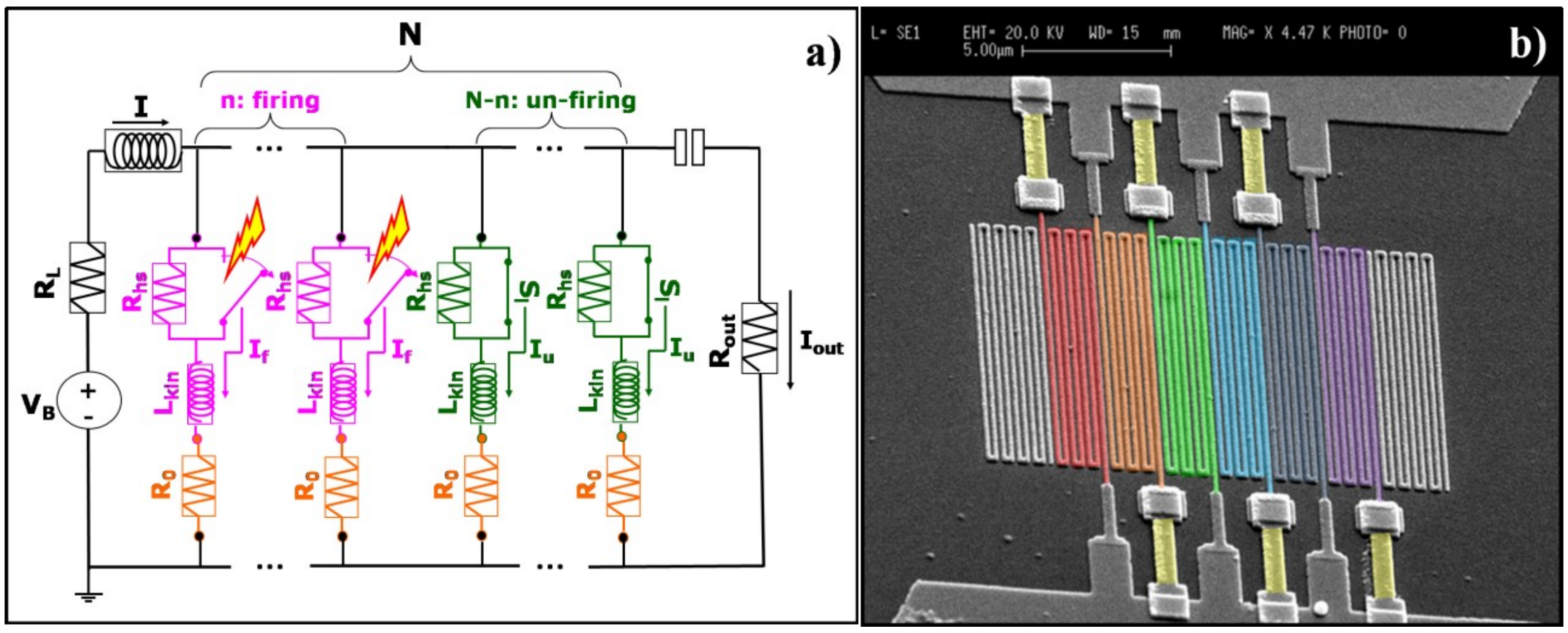

**Figure 1. (a) Circuit equivalent of a N-PND-R. The n firing sections, in pink, all carry the same current $I_f$ and the N-n still superconducting sections (unfiring), in green, all carry the same current $I_u$. $I_{out}$ is the current flowing through the input resistance $R_{out}$ of the preamplifier. (b) Scanning electron microscope (SEM) image of a PND with N=6 and series resistors (6-PND-R) fabricated on a 4nm thick NbN film on MgO. The nanowire width is w=100 nm, the meander fill factor is f=40%. The detector active area is $A_d$=10x10 μm$^2$. The devices are contacted through 70nm thick Au-Ti pads, patterned as a 50 Ω coplanar transmission line. The active nanowires (in color) of the PND-R are connected in series with Au-Pd resistors (in yellow). The floating meanders at the two edges of the PND-R pixel are included to correct for the proximity effect.**

The device shows PNR capability if the height of the current pulse through $R_{out}$ for n firing stripes $\bar{I}_{out}^{(n)}$ is n times higher than the pulse for one $\bar{I}_{out}^{(1)}$, i.e. if the leakage current drained by each of the unfiring nanowires $\delta I_{lk}=I_u-I_B$ is negligible with respect to $I_B$. The leakage current is also undesirable because it lowers the signal available for amplification and temporary increases the current flowing through the still superconducting (unfiring) sections, eventually driving them normal. Consequently, $\delta I_{lk}$ limits the maximum bias current allowed for the stable operation of the device and then the detection efficiencies of the sections. The leakage current depends on the ratio between the impedance of a section $Z_S$ and $R_{out}$ and it can be reduced by engineering the dimensions of the nanowire (thus its kinetic inductance) and of the series resistor (see sec. 6). The design without series resistors simplifies the fabrication process, but, as $Z_S$ is lower, $\delta I_{lk}$ significantly limits the detection efficiency of the device.

## 3. Fabrication

NbN films 3-4 nm thick were grown on MgO <100> substrates (substrate temperature 400°C [23]) by reactive magnetron sputtering in an argon–nitrogen gas mixture. Using an optimized sputtering technique, our NbN samples exhibited a superconducting transition temperature of $T_C$ =10.5 K for 40-Å-thick films. The superconducting transition width was equal to $\Delta T_C$ = 0.3 K.

Both the designs with and without the integrated series resistors were implemented. A scanning-electron microscope (SEM) picture of a 6-PND-R fabricated on MgO is shown in figure 1b.

The size of detector active area ($A_d$) ranges from 5x5 $\mu m^2$ to 10x10 $\mu m^2$ with the number of parallel branches (N) varying from 4 to 14. The nanowires are 100 nm wide and the filling factor (f) of the meander is 40%. The length of each nanowire ranges from 25 to 100 µm.

The three nanolithography steps needed to fabricate the structure have been carried out by using an electron beam lithography (EBL) system equipped with a field emission gun (acceleration voltage 100 kV, 20 nm resolution). In the first step e-beam lithography is used to define pads (patterned as a 50 Ω coplanar transmission line) and alignment markers on a 450nm-thick polymethyl methacrylate (PMMA, a positive tone electronic resist) layer. The sample is then coated with a Ti-Au film (60 nm Au on 10nm Ti) deposited by e-gun evaporation, which is selectively removed by lift-off from un-patterned areas. In the second step, a 160nm thick hydrogen silsesquioxane (HSQ FOX-14, a negative tone electronic resist) mask is defined reproducing the meander pattern. The alignment between the different layers is performed using the markers deposited in the first lithography step. All the unwanted material, i.e. the material not covered by the HSQ mask and the Ti/Au film, is removed by using a fluorine based ($CHF_3$+$SF_6$+Ar ) reactive ion etching (RIE). Finally, with the third step the series resistors (85nm AuPd alloy, 50%-each in weight), aligned with the two previous layers, are fabricated by lift off via a PMMA stencil mask. Our process is optimized to obtain both an excellent alignment between the different e-beam nanolithography steps (error of the order of 100 nm) and a nanowire with high width uniformity (less than 10% [24]).

## 4. Measurement setup

Electrical and optical characterizations have been performed in a cryogenic probe station with an optical window and in a cryogenic dipstick.

In the cryogenic probe station (Janis) the devices were tested at a temperature T=5 K. Electrical contact was realized by a cooled 50 Ω microwave probe attached to a micromanipulator, and connected by a coaxial line to the room-temperature circuitry. The light was fed to the PNDs through a single-mode optical fiber coupled with a long working distance objective, allowing the illumination of a single detector.

In the cryogenic dipstick the devices were tested at 4.2 K. The light was sent through a single-mode optical fiber coupled with a short focal length lens, placed far from the plane of the chip in order to ensure uniform illumination. The number of incident photons per device area was estimated with an error of 5 %.

The bias current was supplied through the DC port of a 10MHz-4GHz bandwidth bias-T connected to a low noise voltage source in series with a bias resistor. The AC port of the bias-T was connected to the room-temperature, low-noise amplifiers. The amplified signal was fed either to a 1 GHz bandwidth single shot oscilloscope or to a 40 GHz bandwidth sampling oscilloscope for time resolved measurements and statistical analysis. The devices were optically tested using a fiber-pigtailed, gain-switched laser diode at 1.3 µm wavelength (100ps-long pulses, repetition rate 26 MHz) and a mode-locked Ti:sapphire laser at 700 nm wavelength (40ps-long pulses, repetition rate 80 MHz).

Throughout the paper, the single photon detection efficiency of an N-PND ($\tilde{\eta}$) or of one of its sections ($\eta$) are defined with respect to the photon flux incident on the area covered by the device (active area $A_d$, typically 10 x 10 $\mu m^2$) or by one section ($A_d/N$), respectively.

## 5. Device characterization

Figure 2a shows a single-shot oscilloscope trace of the photoresponse of a 5-PND under laser illumination (λ=700 nm, 80 MHz repetition rate). Pulses with five different amplitudes can be observed, corresponding to the transition of one to five sections. The measured 80 MHz counting rate represents an improvement of three orders of magnitude over most of the PNR detectors at telecom wavelength [7, 14, 25], with the only exception of the SSPD array [17].

On similar devices, the single-photon detection efficiency ($\tilde{\eta}$) at λ=1.3 μm and the dark-counts rate DK were measured as a function of the bias current at T= 2.2 K [18]. The lowest DK value measured was 0.15 Hz for $\tilde{\eta} \approx 2\%$, yielding a noise equivalent power [26] NEP=$4.2\mathrm{x}10^{-18}$ W/$\mathrm{Hz}^{1/2}$. This sensitivity outperforms most of the other approaches by one-two orders of magnitude (with the only exception of transition-edge sensors [25], which require a much lower operating temperature).

We investigated the temporal response of a 10x10 $\mu m^2$ 4-PND-R probed with light at 1.3 μm wavelength using a 40 GHz sampling oscilloscope (figure 2b). All four possible amplitudes can be observed. The pulses show a full width at half maximum (FWHM) as low as 660ps. In a traditional 10x10 $\mu m^2$ SSPD, the pulse width would be of the order of 10 ns FWHM, so the recovery of the output current $I_{out}$ through the amplifier input resistance is a factor ~$4^2$ faster (see section 6.2), which agrees with results reported by other groups [27, 28]. As shown in section 6.2, the very attractive $N^2$ scaling rule for the output pulse duration unfortunately does not apply to the device recovery time.

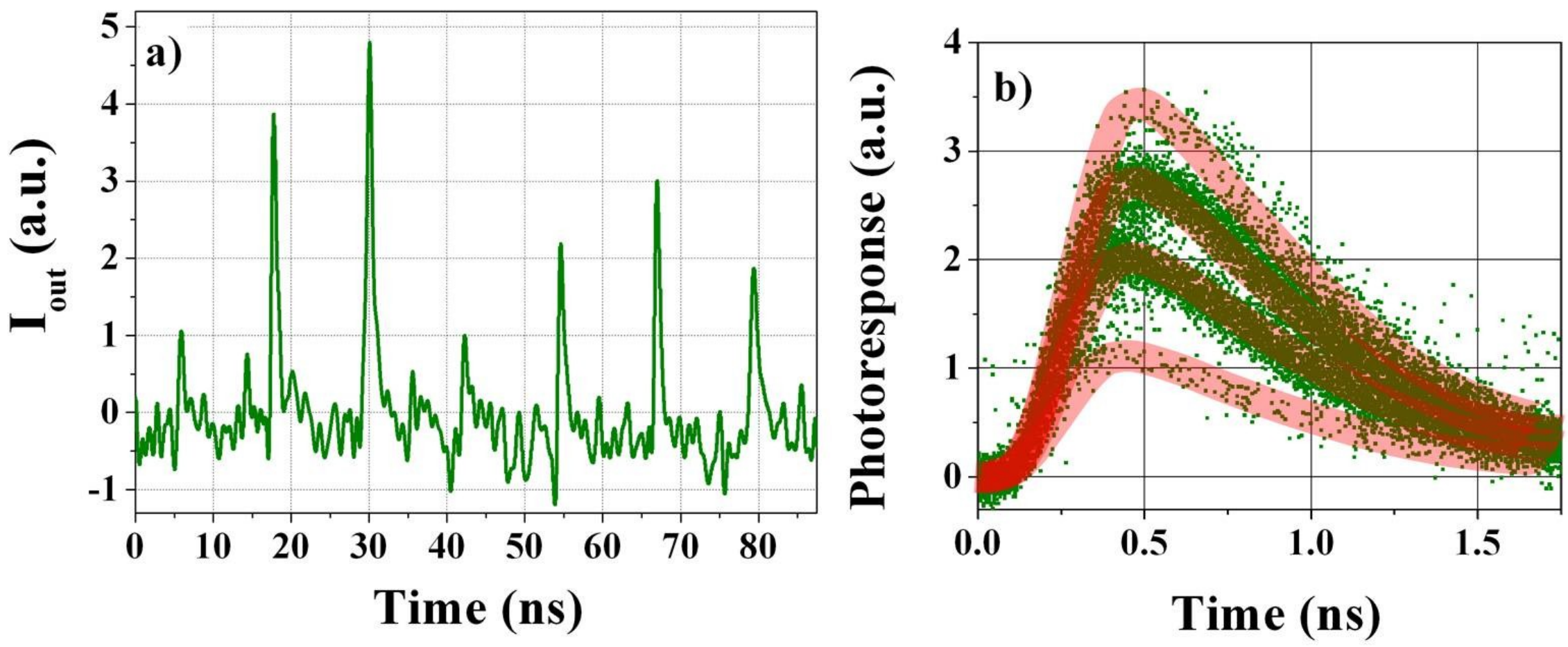

**Figure 2 (a) Single-shot oscilloscope trace during photodetection by a 8.6x8 μm² 5-PND. The device was tested under uniform illumination in a cryogenic dipstick dipped in a liquid He bath at 4.2 K. The light pulses at 700 nm form a mode-locked Ti:sapphire laser had a repetition rate of 80 MHz. (b) Photoresponse transients taken with a 40 GHz sampling oscilloscope while probing a 10x10 μm² 4-PND-R in the cryogenic probe station under illumination with 1.3 μm, 100ps-long pulses from a laser diode, at a repetition rate of 26MHz. The solid curves are guides to the eyes.**

## 6. PND Design

We aim at providing a detailed understanding of the device operation and guidelines for the design of PNDs with optimized performance in terms of efficiency, speed and sensitivity.

The first step is to define the relevant parameter space. The width of the nanowire (w=100 nm) and the filling factor (f=50%) of the meander are fixed by technology, the thickness of the superconducting film (t=4nm) is the optimum value yielding the maximum device efficiency and the active area ($A_d$=10 x 10 $\mu m^2$) is fixed by the size of the core of single mode fibers to which the device must be coupled. We consider single-pass geometries (no optical cavity), but the same guidelines can be applied to cavity devices with optimized absorption [29]. The parameters of the PND-R that can be used as free design variables are: the number of sections in parallel N, the value of the series resistor $R_0$ and the value of the inductance of each section $L_0$. The number of sections in parallel N can be chosen within a discrete set of values (N=2, 3, 4, 6, 7, 10, 17), which satisfy the constraints of w, f, size of the pixel and that the number of stripes in each sections is to be odd (we consider the geometry of Figure 1b). The value of $L_0$ is the sum of the kinetic inductance of each meander $L_{kin}$ and of a series inductance which can be eventually added. $L_{kin}$ is not a design parameter, as it is fixed by w, t, f, $A_d$ and N. If no series inductors are added (bare devices, $L_0=L_{kin}$), the value of $L_0$ for each N is listed in Table 1.

**Table 1. Inductance ($L_0$) and number of squares (SQ) of each section for all possible values of N. The width of the nanowires is w=100 nm, the thickness is t=4 nm. The kinetic inductance per square was estimated ($L_{kin}/\square=90$ pH) from the time constant of the exponential decay of the output current ($\tau_{out}=\tau_f=L_{kin}/R_{out}$, see sec. 6.2) for a standard 5x5μm$^2$ SSPD [23].**

| N | $L_0$ | SQ |
|---|---|---|
| 2 | 225 nH | 2500 |
| 3 | 153 nH | 1700 |
| 4 | 117 nH | 1300 |
| 6 | 81 nH | 900 |
| 7 | 63 nH | 700 |
| 10 | 45 nH | 500 |
| 17 | 27 nH | 300 |

An additional free parameter, relative to the read-out, is the impedance seen by the device on the RF section of the circuit $R_{out}$, which is 50 Ω (of the matched transmission line) in the actual measurement setup (see section 4), but which can be varied in principle from zero to infinite introducing a cold preamplifier stage.

The target performance specifications are the single-photon detection efficiency (η), the signal to noise ratio (SNR) and the maximum repetition rate (speed), which must be optimized under the constraints that the operation of the device is stable and that it is possible to detect a certain maximum number of photons ($n_{max}$) dependent on the specific application.

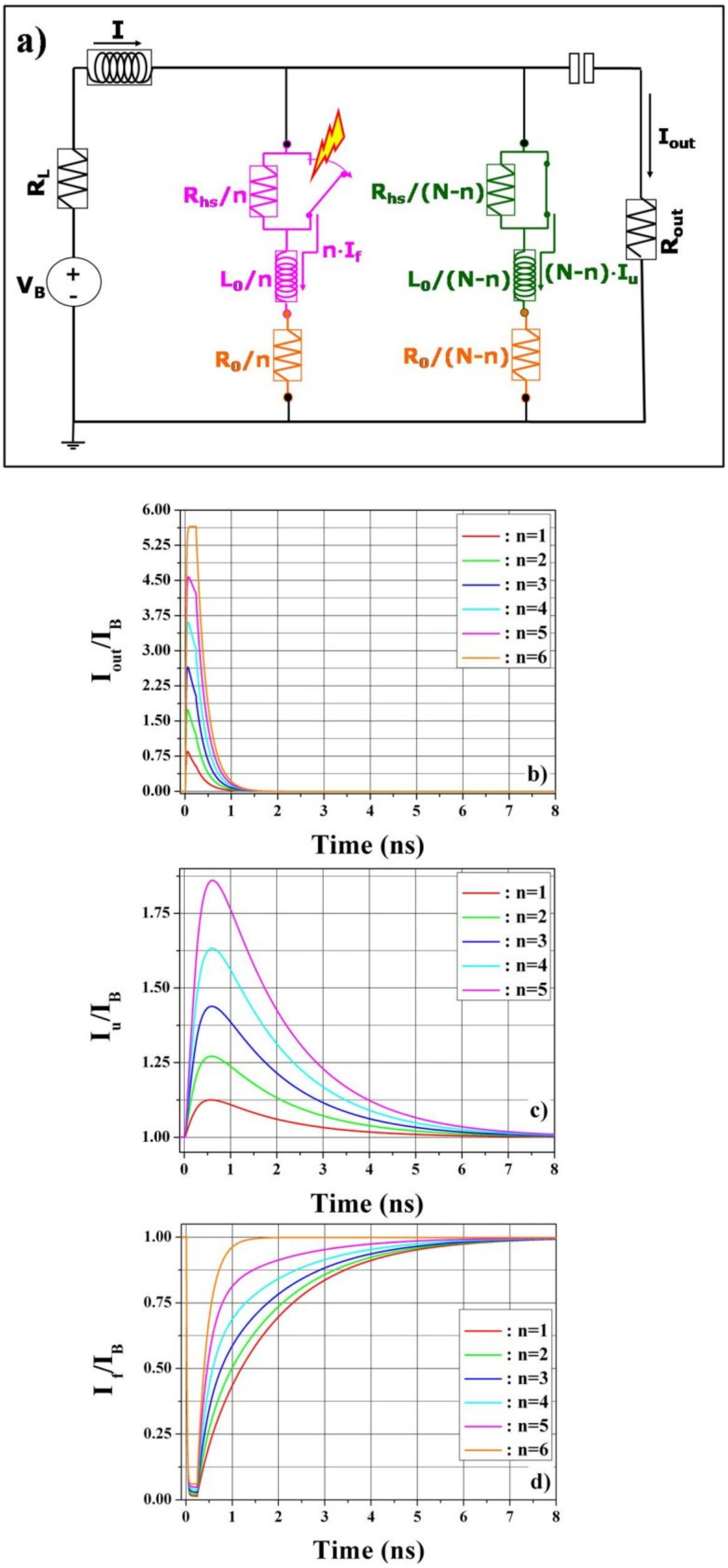

**Figure 3. (a) Simplified circuit of a N-PND-R, where the two sets of n firing and the N-n unfiring sections have been substituted by their Thévenin-equivalents. (b-d) Simulated time evolution of $I_u$ (b), $I_{out}$ (c) and $I_f$ (d) for a 6-PND-R as n increases from 1 to 6. The parameters of the circuit are: $L_0=L_{kin}=81$ nH, $R_0=50\ \Omega$, $R_{out}=50\ \Omega$, $R_{hs}=5.5$ kΩ, and $t_{hs}=250$ps.**

A comprehensive description of PND operation should combine thermal and electrical modeling of the nanowires [30]. In this work, a purely electrical model (see section 2 and Figure 1a) has been used to make a reliable guess on how the performance of the device varies when moving in the parameter space.
In this model, the dependence of $L_{kin}$ on the current flowing through the nanowire was disregarded, and it was assumed constant. Furthermore, it has been shown [30] that changing the values of the kinetic inductance of an SSPD or of a resistor connected in series to it results in a change of the hotspot resistance and of its lifetime, eventually causing the device to latch to the normal state. The simplified analysis presented here does not take into account these effects, and considers both $R_{hs}$ and $t_{hs}$ as constant ($R_{hs}$=5.5 kΩ, $t_{hs}$=250ps), and that device cannot latch. However, the results of this approach can still quantitatively predict the behavior of the device in the limit where the fastest time constant of the circuit $\tau_f$ (see section 6.2) is much higher than the hotspot lifetime ($\tau_f >> t_{hs}$), and give a reasonable qualitative understanding of the main trends of variation of the performance of faster devices ($\tau_f \sim t_{hs}$).
In order to gain a better insight on the circuit dynamics (see sec. 6.2) and to reduce the calculation time, the N+1 mesh circuit of Figure 1a can be simplified to the three mesh circuit of Figure 3a applying the Thévenin theorem on the n firing sections and on the remaining N-n still superconducting (unfiring) sections, separately.
Figure 3b to d show the simulation results for the time evolution of the currents flowing through $R_{out}$ and through the unfiring ($I_u$) and firing ($I_f$) sections of a PND with 6 sections and integrated resistors (6-PND-R) and for the number of firing sections n ranging from 1 to 6. As n increases, the peak values of the output current ($I_{out,}$ figure 3b) and of the current through the unfiring sections ($I_u$, figure 3c) increase. The firing sections experience a large drop in their current ($I_f$, figure 3d), which is roughly independent on n. The observed temporal dynamics will be examined in the following sections.

### *6.1. Current redistribution and efficiency*

Let $\overline{\delta I}_{lk}^{(n)}$ be the peak value of the leakage current drained by each of the still superconducting (unfiring) nanowires when n sections fire simultaneously. The stability requirement translates in the condition that for each unfiring section $I_B + \overline{\delta I}_{lk}^{(n_{max})} \leq I_C$ (as the leakage current increases with n, $\overline{\delta I}_{lk}^{(n_{max})}$ represents the worst case). This limits the bias current and therefore the single-photon detection efficiency (η), which, for a certain nanowire geometry (i.e. w, t fixed), is a monotonically increasing function of $I_B/I_C$ [20]. For instance, in order to detect a single photon (at λ=1.3 μm, T=1.8K) in a section with an efficiency equal to 80% of the maximum value set by absorption (~32%, [27]), $\overline{\delta I}_{lk}^{(n_{max})}$ should be made ≤33% of $I_B$. Therefore the leakage current strongly affects the performance of the device and it is to be minimized, which makes it very important to understand its dependency from the design parameters: $\overline{\delta I}_{lk}^{(n)}\left(N, L_0, R_0, R_{out}\right)$.

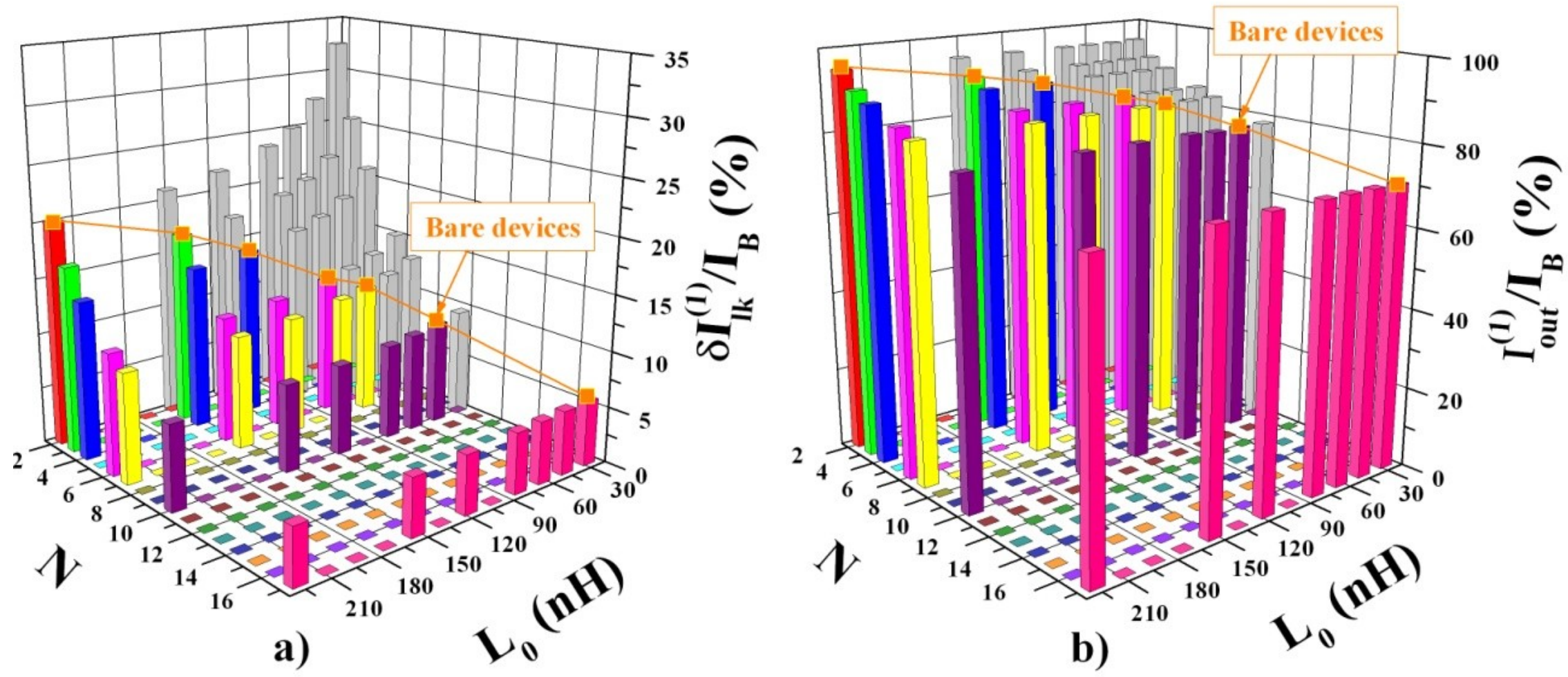

**Figure 4. Peak value of the leakage current $\overline{\delta I}_{lk}^{(1)}$ drained by each of the still superconducting (unfiring) nanowires (a) and of the output current $\overline{I}_{out}^{(1)}$ (b) when only one section fires plotted as a function of the number of sections in parallel N and of the value of the inductance of each section $L_0$. The value of the series resistor $R_0$ and of the output resistor $R_{out}$ is 50 Ω. The orange line highlights bare devices, the colored bars correspond to devices which respect the constraints on the geometry of the structure while the grey bars refer to purely theoretical devices which just show the general trend. The leakage current and the output current are expressed in % of the bias current $I_B$ because they are proportional to it.**

The leakage current for n=1 is first investigated and its dependency on n is then presented for some particular combinations of design parameters. The dependency of $\overline{\delta I}_{lk}^{(1)}$ on N and $L_0$ at fixed $R_0$ and $R_{out}$ (both equal to 50 Ω) is shown in figure 4a: an orange line highlights bare devices ($L_0$=$L_{kin}$, see Table 1) and the colored bars are relative to devices which respect the constraints on the geometry of the structure ($L_0$>$L_{kin}$), while the grey bars refer to purely theoretical devices which just show the general trend. For any N, the current redistribution increases with decreasing $L_0$, as the impedance of each section decreases. Keeping $L_0$ constant, $\overline{\delta I}_{lk}^{(1)}$ decreases with increasing N, as the current to be redistributed is fixed and the number of channels draining current increases. For this reason also the increase of redistribution with decreasing $L_0$ becomes weaker for high N.

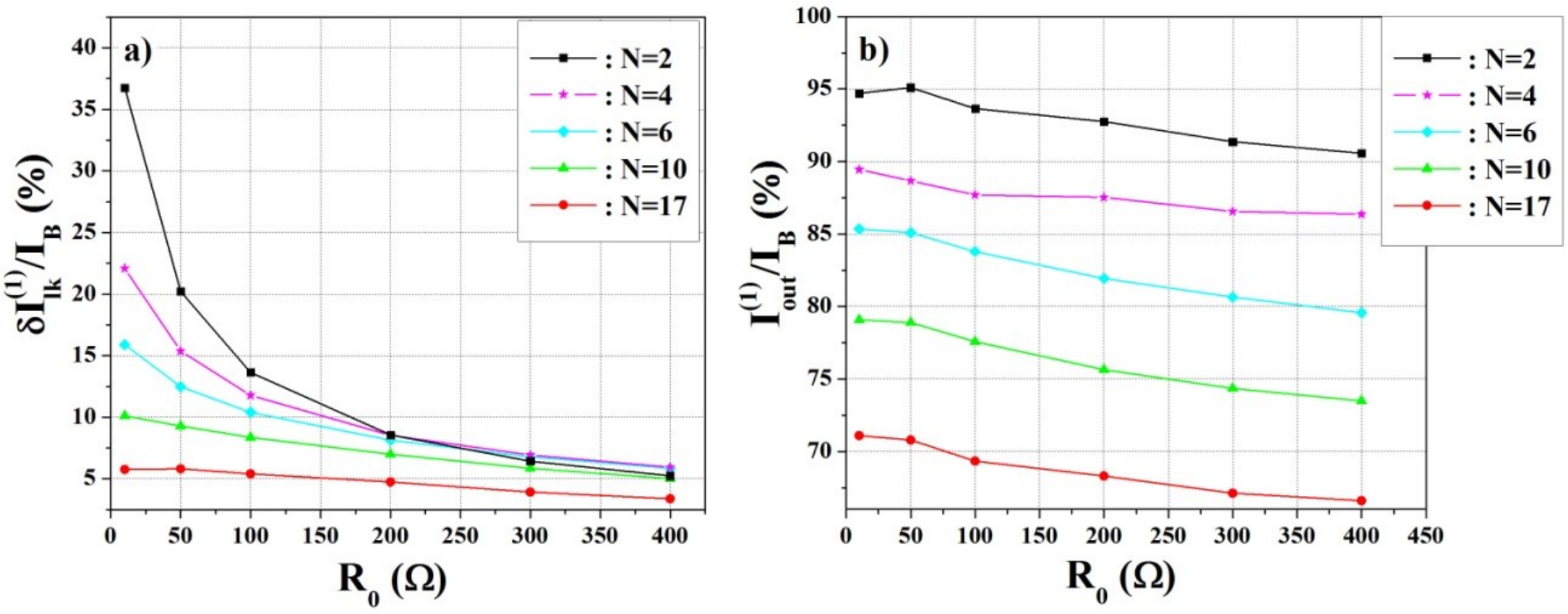

**Figure 5. (a) Variation of the peak value of the leakage current per unfiring section of some bare devices for n=1 ($\overline{\delta I}_{lk}^{(1)}$) as the resistance of the series resistor $R_0$ varies from 10 Ω to 400 Ω. (b) Peak value of the output current for n=1 ($\overline{I}_{out}^{(1)}$) as a function of $R_0$ for some bare devices.**

The dependency of $\overline{\delta I}_{lk}^{(1)}$ on $R_0$ is shown in figure 5a for some bare devices and $R_{out}$=50 Ω. As expected, the redistribution decreases as $R_0$ increases because the impedance of each section increases with respect to the output resistance. For the same reason, $\overline{\delta I}_{lk}^{(1)}$ is strongly reduced (to ~3% of $I_B$) when $R_{out}$ is decreased of one order of magnitude from 50 to 5 Ω, keeping $R_0$ constant (figure 6a).

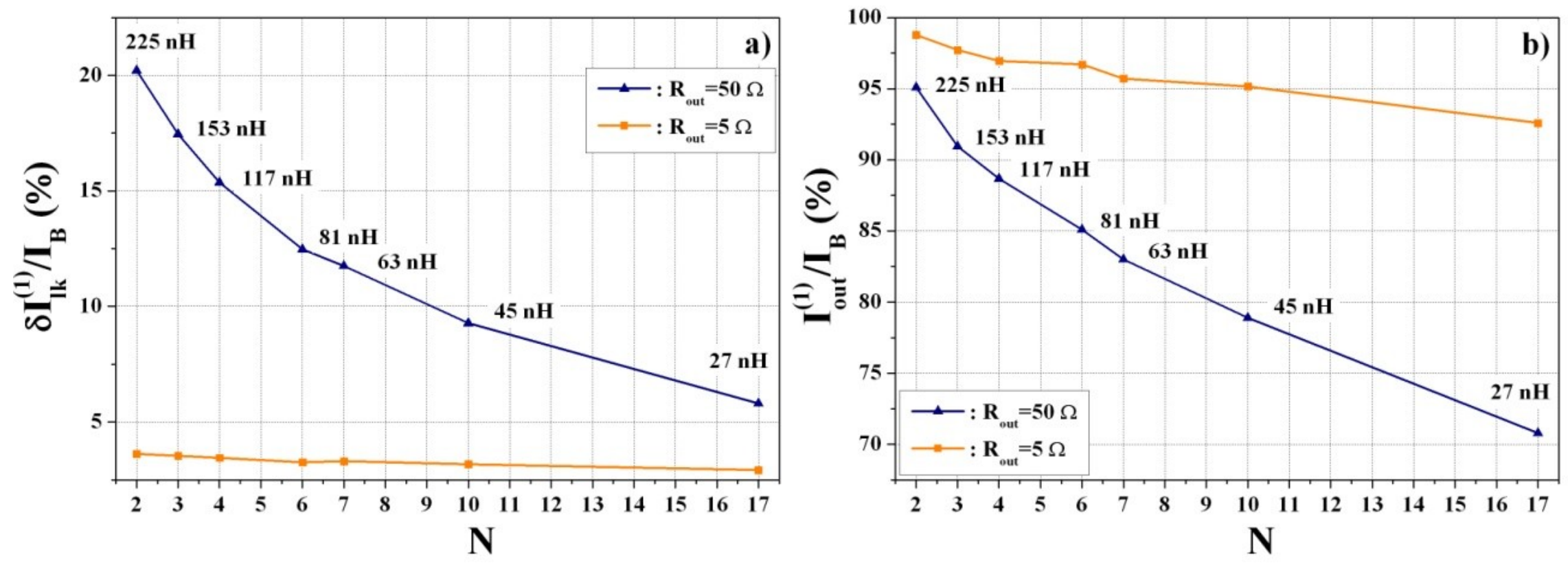

**Figure 6. Variation of the peak value of the leakage current per unfiring section (a) and of the output current (b) of the set of bare devices for n=1 ($\overline{\delta I}_{lk}^{(1)}$ and $\overline{I}_{out}^{(1)}$, respectively) as the resistance of the output resistor $R_{out}$ decreases of one order of magnitude from 50 to 5 Ω (in blue and orange, respectively), while $R_0$=50 Ω.**

The variation of the leakage current with the number of firing stripes n ($\overline{\delta I}_{lk}^{(n)}$) for the set of bare devices is presented in figure 7a. The dependency is superlinear ($\overline{\delta I}_{lk}^{(n)} > n\overline{\delta I}_{lk}^{(1)}$), as the current to be redistributed per firing stripe is always the same (see sec. 6.3), but the number of channels draining current decreases. Furthermore, as expected, the curves for different design parameter sets never cross, which means that all the design guidelines presented in Figure 5a, Figure 6a, Figure 7a for n=1 still apply for higher n.

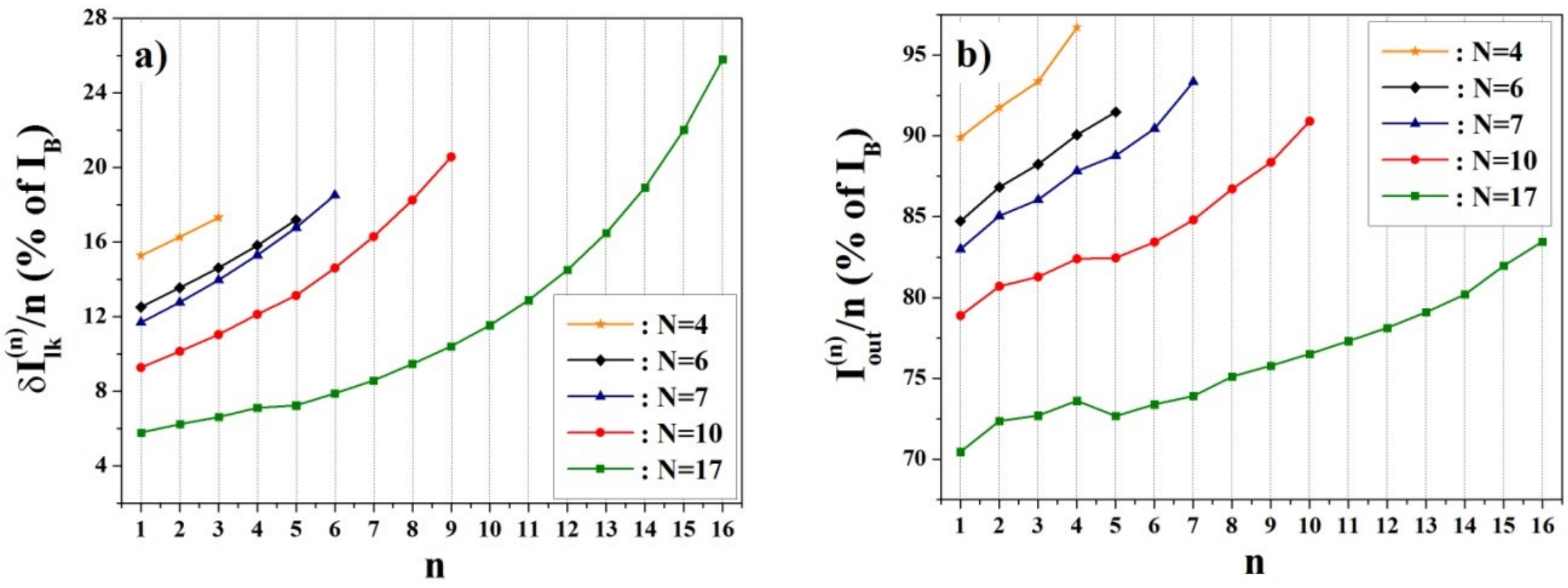

**Figure 7. Variation of the leakage current $\overline{\delta I_{lk}}^{(n)}$ (a) and of the output current $\overline{I_{out}}^{(n)}$ (b) with the number n of firing stripes for the set of bare devices.**

In conclusion, the result of this simplified analysis is that, in order to minimize the leakage current and thus maximize the efficiency, N, $L_0$ and $R_0$ must be made as high as possible and $R_{out}$ as low as possible. We note however that $R_0$ cannot be increased indefinitely to avoid that the nanowire latches to the hotspot plateau before $I_B$ reaches $I_C$ [23].

### *6.2. Transient response and speed*

Before proceeding to the analysis of the SNR and speed performances of the device, it is necessary to discuss the characteristic recovery times of the currents in the circuit.

The transient response of the simplified equivalent electrical circuit of the N-PND (figure 3a) to an excitation produced in the firing branch can be easily found analytically. Therefore, the transient response of the current through the firing sections $I_f$, through the unfiring sections $I_u$ and through the output $I_{out}$ after the nanowires become superconducting again ($t \geq t_{hs}$) can be written as:

$$\begin{cases} I_f \propto \frac{N-n}{N}\exp\left(-t/\tau_s\right)+\frac{n}{N}\exp\left(-t/\tau_f\right) \\ I_u \propto \exp\left(-t/\tau_s\right)-\exp\left(-t/\tau_f\right) \\ I_{out} \propto \exp\left(-t/\tau_f\right) \end{cases} \qquad (1)$$

where $\tau_s=L_0/R_0$ and $\tau_f=L_0/(R_0+NR_{out})$ are the "slow" and the "fast" time constant of the circuit, respectively.

This set of equations describes quantitatively the time evolution of the currents after the healing of the hotspot in the case $\tau_f >> t_{hs}$, and it provides a qualitative understanding of the recovery dynamics of the circuit for shorter $\tau_f$.

The recovery transients ($t \geq t_{hs}$) of $I_{out}$, $\delta I_{lk}$ and $I_f$ for a 4-PND-R simulated with the circuit of figure 3a are shown in figure 8a, b, c, respectively (in blue) for different number of firing sections (n=1 to 4). As n increases from 1 to 4, the recoveries of $I_{out}$ and $\delta I_{lk}$ change only by a scale factor. On the other hand, the transient of $I_f$ depends on n and becomes faster increasing n, as qualitatively predicted by the first of equations (1). Indeed, $I_f$ consists in the sum of a slow and a fast contribution, whose balance is controlled by the number of firing sections n. To prove the quantitative agreement with the analytical model in the limit $\tau_f >> t_{hs}$, the simulated transients of $I_{out}$, $\delta I_{lk}$ and $I_f$ have been fitted (figure 8a, b, c, respectively, in orange) using the set of equations (1), and four fitting parameters ($\tau_s$, $\tau_f$, a time offset $t_0$ and a scaling factor K). The values of $\tau_s$ and $\tau_f$ obtained from the three fittings (of $I_{out}$, of $\delta I_{lk}$ and of the whole set of four $I_f$ for n=1,…, 4) closely agree with the values calculated from the analytical expressions presented above and the parameters of the circuit ($\tau_s^*$=2.30 ns ,$\tau_f^*$=460 ps).

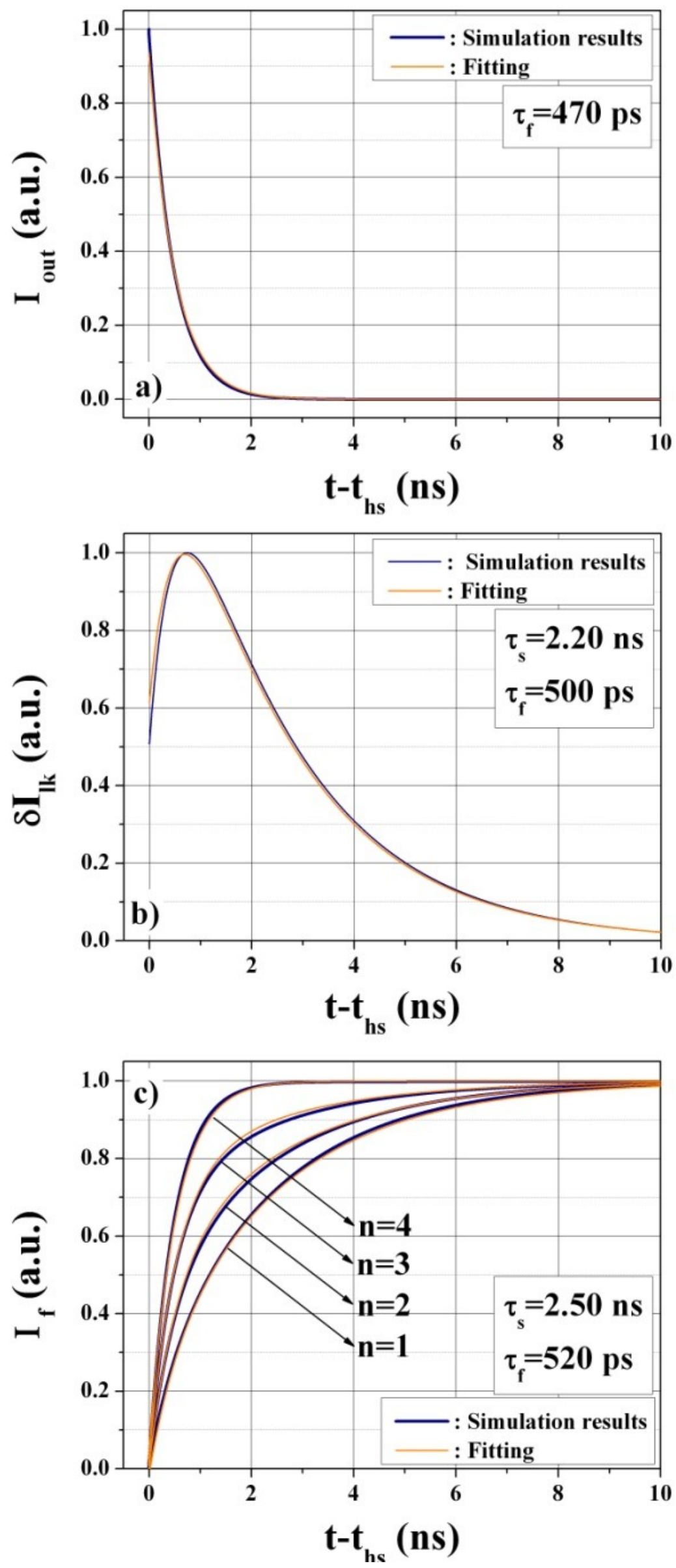

**Figure 8. Recovery transients ($t \geq t_{hs}$) of $I_{out}$ (a), $\delta I_{lk}$ (b), and $I_f$ (c) for a 4-PND-R as n increases from 1 to 4. The simulated transients are in blue, the fitted curves are in orange. The parameters of the circuit used for the simulations are: $L_0=L_{kin}$=117 nH, $R_0$=50 Ω, $R_{out}$=50 Ω, $R_{hs}$=5.5 kΩ, and $t_{hs}$=250ps. The three sets of curves are fitted by equations (1) (multiplied by K, and shifted by $t_0$), where the values of $\tau_s$ and $\tau_f$ are shown in the insets.**

In order to quantify the speed of the device, we take $f_0=(t_{reset})^{-1}$ as the maximum repetition frequency, where $t_{reset}$ is the time that $I_f$ needs to recover to 95% of the bias current after a detection event.

According to the results presented above, which are in good agreement with experimental data (figure 2b), $I_{out}$ decays exponentially with the same time constant for any n ($\tau_{out}=\tau_f$), which, for a bare N-PND, is $N^2$ times shorter than a normal SSPD of the same surface [27, 28]. This however does not relate with the speed of the device. Indeed, $t_{reset}$ is the time that the current through the firing sections $I_f$ needs to rise back to its steady-state value ($I_f \sim I_B$). In the best case of n=N, $I_f$ rises with the fast time constant $\tau_f$, but in all other cases the slow contribution becomes more important as n decreases (see figure 3d and figure 8c), until, for n=1, $I_f \sim [1-\exp(-t/\tau_s)]$. The speed performance of the device is then limited by the slow time constant ($t_{reset} \sim 3 \cdot \tau_s$), which means that an N-PND is only N times faster than a normal SSPD of the same surface, being as fast as a normal SSPD whose kinetic inductance is the same as one of the N section of the N-PND.

Figure 9 shows the dependence of $f_0$ on $L_0$ and $R_0$. For $\tau_s < t_{hs}$ (i.e. $f_0 > 4$ GHz in our model) the speed of the device may be limited by the hotspot temporal dynamics, and so no reliable predictions can be made using our simplified model.

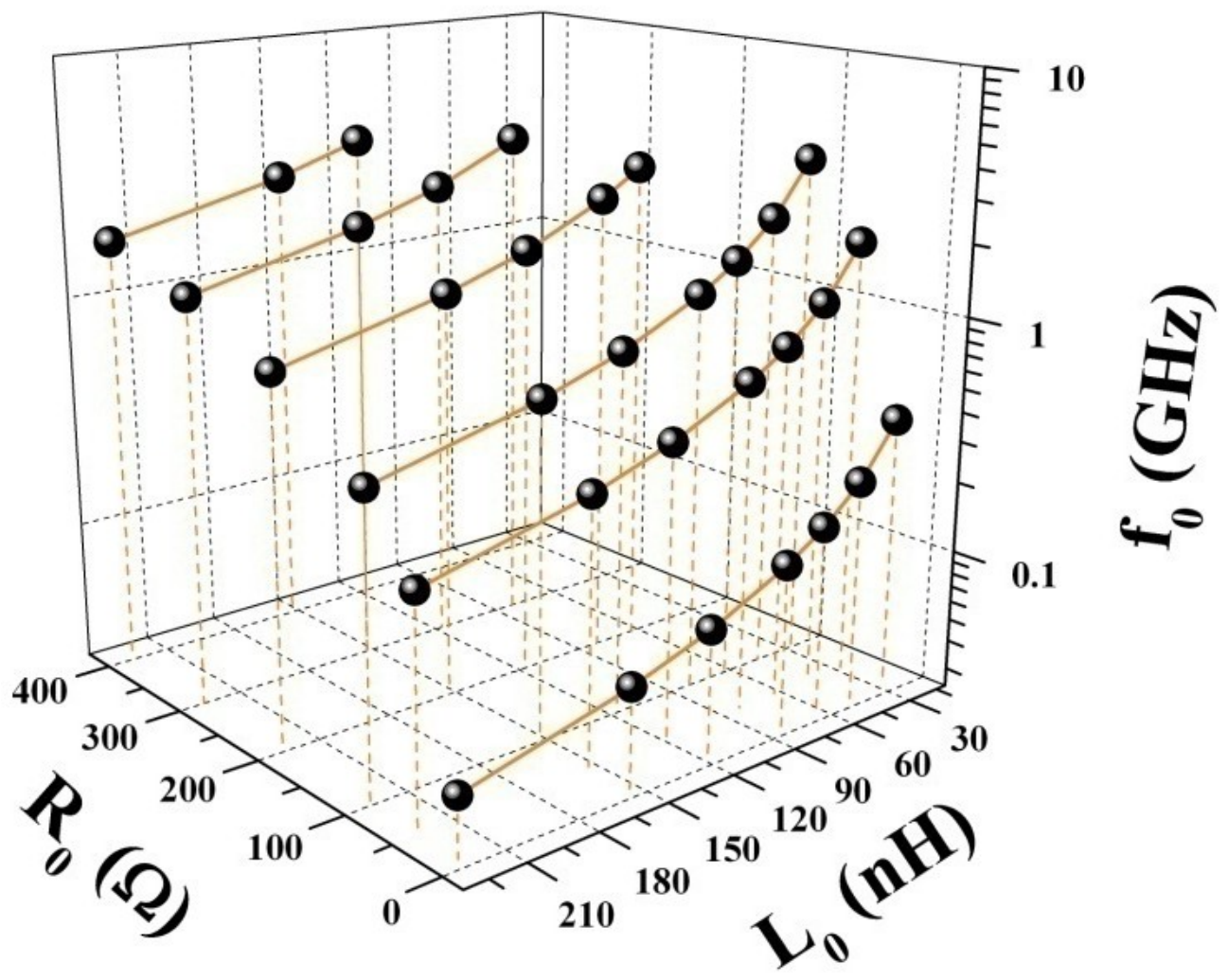

**Figure 9. Dependence of $f_0$ on $L_0$ and $R_0$. No data are presented for $f_0$>4 GHz, where no reliable predictions can be made using this simplified model.**

### *6.3. Signal to noise ratio*

The peak value and the duration of the output current pulse are a function of the design parameters (see below and section 6.2, respectively). As the output pulse becomes faster, amplifiers with larger bandwidth are required and thus electrical noise become more important. In order to assess the possibility to discriminate the output pulse from the noise, we define the signal to noise ratio (SNR) as the ratio between the maximum of the output current $\bar{I}_{out}$ and the rms value of the noise-current at the preamplifier input $I_n$, $SNR = \bar{I}_{out}/I_n$.

The peak value of the output current when n sections fire simultaneously (see figure 3b, relative to a 6-PND-R) can be written as:

$$I_{out}^{(n)} = n\left(I_B - I_f^{(n)*}\right) - \left(N-n\right)\delta I_{lk}^{(n)*}$$

where the starred values refer to the time $t=t^*$ when the output current peaks.

As n=1 represents the worst case, in order to evaluate the performance of the device in terms of the SNR, the dependency of $\bar{I}_{out}^{(1)}$ from the design parameters is investigated: $\bar{I}_{out}^{(1)}\left(N, L_0, R_0, R_{out}\right)$. The dependency of $\bar{I}_{out}^{(1)}$ on N and $L_0$ at fixed $R_0$ and $R_{out}$ (both equal to 50 Ω) is shown in figure 4b. Inspecting the values of $\bar{I}_{out}^{(1)}$ and of $\overline{\delta I}_{lk}^{(1)}$ for the same device in figure 4, it becomes clear that they add up to a value well above to $I_B$, which is due to the fact that the output current and of the leakage current peak at two different times, t* and $t_{lk}$, respectively (figure 3). Furthermore, as $t_{lk}>t^*$, the output current is not significantly affected by redistribution, because $I_{out}$ is maximum when $\delta I_{lk}$ is still beginning to rise.
The expression for $t_{lk}$ can be derived from eq. (1): $t_{lk}=L_0/(N \cdot R_{out})\ln(1+N \cdot R_{out}/R_0)$, which means that increasing the device speed (decreasing $L_0$ or $R_0$), N or $R_{out}$ makes the redistribution faster and then $\bar{I}_{out}^{(1)}$ lower.

So, for any given N, $\bar{I}_{out}^{(1)}$ decreases (figure 4b) with decreasing $L_0$, both because $\overline{\delta I}_{lk}^{(1)}$ is higher and because $t_{lk}$ is lower. Keeping $L_0$ constant, $\bar{I}_{out}^{(1)}$ decreases with increasing N because even though $\overline{\delta I}_{lk}^{(1)}$ decreases, the redistribution peaks earlier and the number of channels draining current increases.

The dependency of $\bar{I}_{out}^{(1)}$ on $R_0$ is shown in figure 5b for some bare devices and $R_{out}$=50 Ω. Even though $\overline{\delta I}_{lk}^{(1)}$ decreases as $R_0$ increases (figure 5a), the output current decreases due to the redistribution speed-up (decrease of $t_{lk}$): $\delta I_{lk}^{(1)*}$ increases despite of the decrease of the peak value of the leakage current. On the other hand, a decrease in $R_{out}$ makes the redistribution much less effective, as $t_{lk}$ decreases slower with decreasing $R_{out}$ than with increasing $R_0$. Indeed, as shown in figure 6b for bare devices, $\bar{I}_{out}^{(1)}$ significantly increases when $R_{out}$ is decreased by one order of magnitude from 50 to 5 Ω, keeping $R_0$ constant.
In conclusion, in order to maximize the output current, N, $R_0$ and $R_{out}$ must be minimized, while $L_0$ must be made as high as possible.

The rms value of noise-current at the preamplifier input $I_n$ can be written as $I_n=\sqrt{S_n \Delta f}$, where $S_n$ is the noise spectral power density of the preamplifier and Δf is the bandwidth of the output current $I_{out}$, which is estimated as $\Delta f=1/\tau_{out}$, where $\tau_{out}=\tau_f=L_0/(R_0+NR_{out})$ is the time constant of the exponential decay of $I_{out}$ (see sec. 6.2). $I_n$ is then a function of the parameters of the device and of the read-out through $S_n$ and $\tau_f$. As $S_n$ varies consistently with the type of preamplifier used, in our analysis of $I_n$ we take into account only the variation of Δf. Assuming $S_n$ constant, $I_n$ is thus minimized minimizing N, $R_0$ and $R_{out}$ and maximizing $L_0$.

The same optimization criteria apply then naturally to the SNR. The dependence of $\bar{I}_{out}^{(1)}/\sqrt{\Delta f}$ on N and $L_0$ for an input resistance of Rout=50Ω is shown in figure 10.

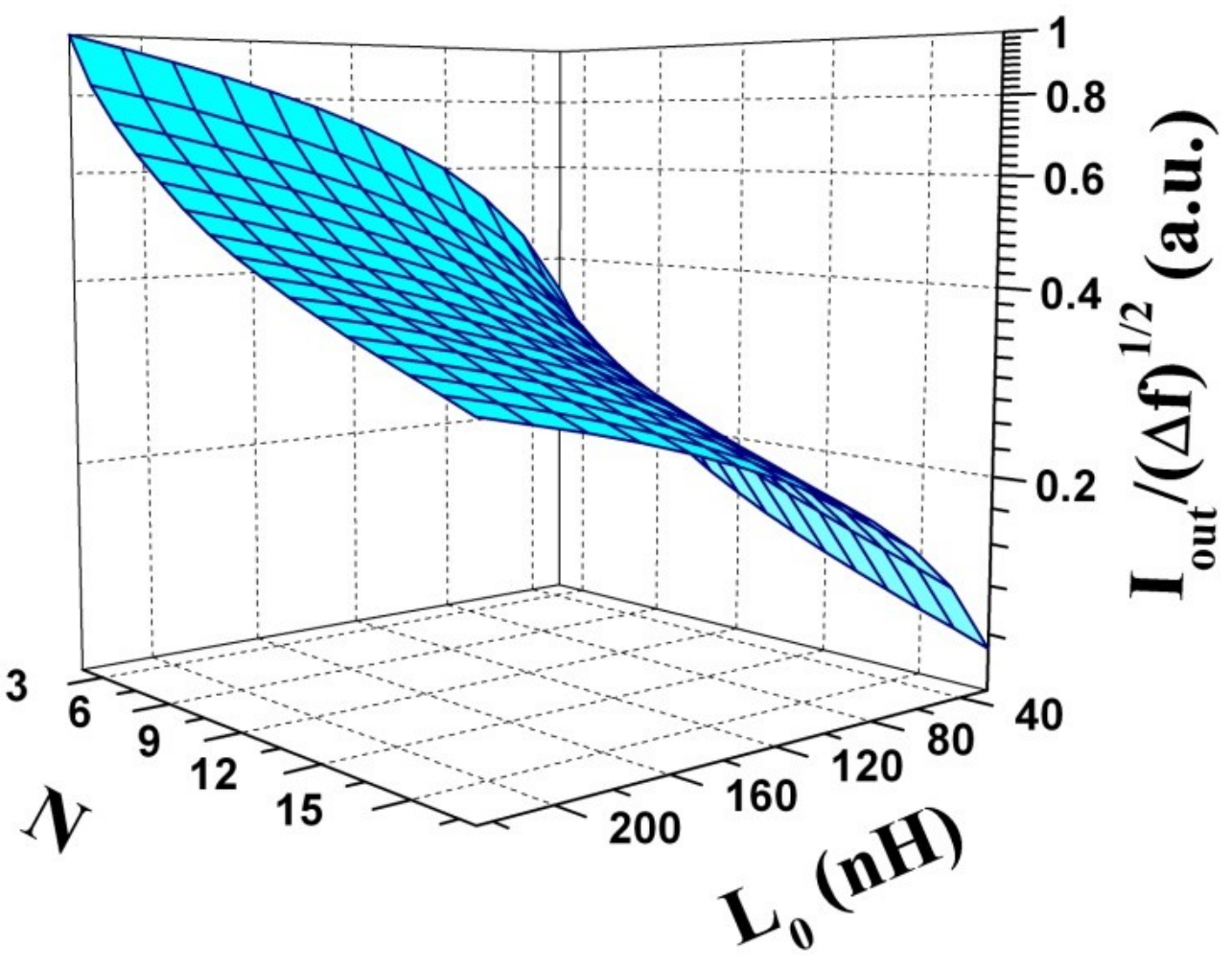

**Figure 10.** $\bar{I}_{out}^{(1)}/\sqrt{\Delta f}$ **as a function of N and $L_0$.**

The main design guidelines which can be deduced from the analysis of sections 6.1 to 6.3 are summarized in Table 2. The type of dependency of $\overline{\delta I}_{lk}$, $f_0$, $\bar{I}_{out}$ and $\Delta f$ from the design parameters ($L_0$, $R_0$, $R_{out}$, N) is indicated.

**Table 2. Dependency of $\overline{\delta I}_{lk}$, $f_0$, $\bar{I}_{out}$ and $\Delta f$ from the design parameters: increasing with increasing the parameter (↗), decreasing with increasing the parameter (↘), independent (––).**

| | $L_0$ | $R_0$ | $R_{out}$ | N |
|---|---|---|---|---|
| $\delta I_{lk}$ | ↘ | ↘ | ↗ | ↘ |
| $f_0$ | ↘ | ↗ | –– | –– |
| $I_{out}$ | ↗ | ↘ | ↘ | ↘ |
| $\Delta f$ | ↘ | ↗ | ↗ | ↗ |

## 7. Application to the measurement of photon number statistics

We wish to determine whether the PND can be used to measure an unknown photon number probability distribution. Indeed, the light statistics measured with a PND differ from the original one due to non-idealities such as the limited number of sections and limited and non-uniform efficiencies ($\eta_i$) of the different sections. As a PND can be modeled as a balanced lossless N-port beam splitter, every channel terminating with a single photon detector [19], the input-output transformation can be formalized as follows.

Let an N-PND be probed with a light whose photon number probability distribution is $\mathbf{S}=[S(m)]=[s_m]$, and its output be sampled H times. The result of the observation can be of N+1 different types (i.e. 0,..., N stripes firing), so an histogram of the H events can be constructed, which can be represented by a (N+1)-dimensional vector $\mathbf{r}=[r_i]$, where $r_i$ is the number of runs in which the outcome was of the $i^{th}$ type. The expectation value of the statistics obtained from the histogram is $E[\mathbf{Q}_{ex}=\mathbf{r}/H]=\mathbf{Q}$, where $\mathbf{Q}=[Q(n)]=[q_n]$ is the probability distribution of the number of measured photons.

Q(n) is related to the incoming distribution S(m) by the relation:

$$Q(n) = \sum_m P^N(n \mid m) \cdot S(m) \qquad (2)$$

where $P^N(n|m)$ is the probability that n photons are detected when m are sent to the device. Eq. (2) may be rewritten in a matrix form as $\mathbf{Q}=\mathbf{P^N}\cdot\mathbf{S}$, where $P^N=\left[P^N(n|m)\right]=\left[p^N_{nm}\right]$ is the matrix of the conditional probabilities.

It has been shown [31, 32] that an unknown incoming photon number distribution **S** can be recovered if **Q** and $\mathbf{P^N}$ are known.

### *7.1. Matrix of conditional probabilities*

The matrix of the conditional probabilities of a N-PND depends only on the vector of the N single-photon detection efficiencies of the different sections of the device $\boldsymbol{\eta}=[\eta_i]$ through the relations presented in [18, 19]. The vector $\boldsymbol{\eta}$ can be then determined from the statistics $\mathbf{Q_{ex}}$ measured when probing the device with a light of known statistics **S**.

For example, using a laser light source with Poissonian photon number probability distribution, the probability distribution of the number of measured photons **Q** (expressed by (2)) was fitted to the experimentally measured distribution $\mathbf{Q_{ex}}$ using $\boldsymbol{\eta}$ as a free parameter. The resulting $\boldsymbol{\eta}$ and matrix of the conditional probabilities are shown in figure 11 for a 5-PND at λ=700 nm.

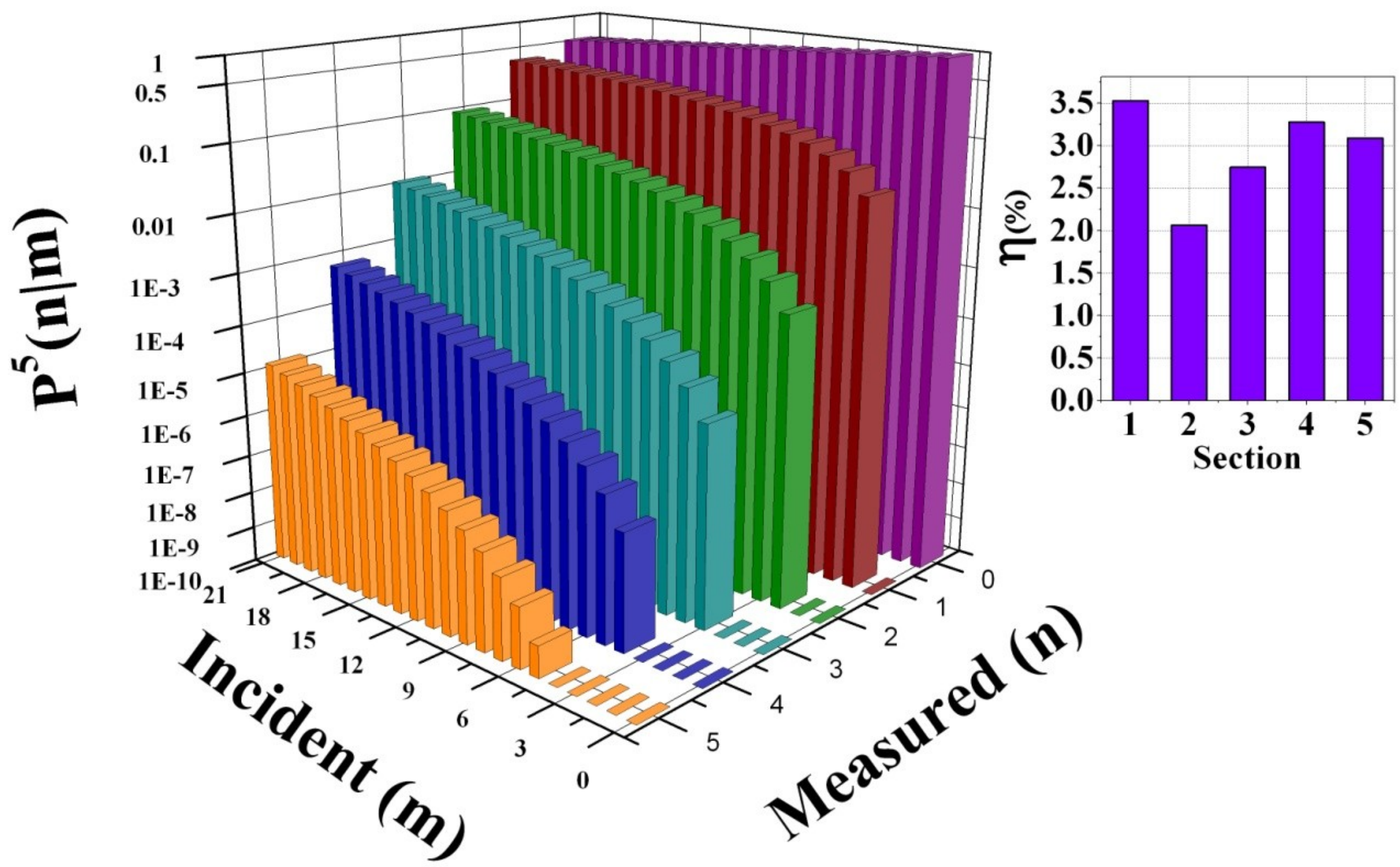

**Figure 11. Conditional probability matrix for a 8.6x8 µm² 5-PND (with no integrated series resistors), calculated from the vector η of the 5 single-photon detection efficiencies (relative to T=4.2 K, λ=700nm) of the different sections of the device (inset).**

The $\mathbf{P^N}$ matrix provides a full description of the detector. Once $\mathbf{P^N}$ is known, several approaches can be used to reconstruct **S** from the histogram **r**. In the case no assumptions on the form of **S** are made, the maximum likelihood (ML) method is the most suitable, as it is the most efficient in solving this class of problems **[33]**.

### *7.2. Maximum-Likelihood (ML) method*

Let $\mathbf{R}=R_0,\ldots, R_N$ be the random vector of the populations of the (N+1) different bins of the histogram after H observations. The joint probability density function $L(\mathbf{r}|\mathbf{Q})$ for the occurrence of the particular configuration $\mathbf{r}=r_0,\ldots, r_N$ of $\mathbf{R}$ is called the likelihood function of $\mathbf{r}$ and it is given by [34]:

$$L(\mathbf{r}|\mathbf{Q}) = H!\prod_{i=0}^{N}\frac{q_i^{r_i}}{r_i!} \qquad (3)$$

where $\mathbf{Q}=[q_i]$ is the probability distribution of the number of measured photons, i.e. the vector of the probabilities to have an outcome in the bin i (i=0,…, N) in a single trial.
Considering $\mathbf{Q}$ as a function of $\mathbf{S}$ through equation (2), we can rewrite the likelihood function of the vector $\mathbf{r}$, depending on the parameter $\mathbf{S}$:

$$L(\mathbf{r}|\mathbf{S}) = H!\prod_{i=0}^{N}\frac{\left(\sum_m p_{i,m}^N s_m\right)^{r_i}}{r_i!} \qquad (4)$$

which is then the probability of the occurrence of the particular histogram $\mathbf{r}$ when the incoming light has a certain statistics $\mathbf{S}$.
As $\mathbf{r}$ is measured and then it is known, $L(\mathbf{r}|\mathbf{S})$ can be regarded as a function of $\mathbf{S}$ only, i.e. $L(\mathbf{r}|\mathbf{S})$ is the probability that a certain vector $\mathbf{S}$ is the incoming probability distribution when the histogram $\mathbf{r}$ is measured. The best estimate of the incoming statistics which produced the histogram $\mathbf{r}$ according to the ML method is the vector $\mathbf{S_e}$ which maximizes $L(\mathbf{r}|\mathbf{S})$, where $\mathbf{r}$ is treated as fixed. So, the estimation problem can in the end be reduced to a maximization problem.

For numerical calculations, it is necessary to limit the maximum number of incoming photons to $m_{max}$ (in the following calculations, $m_{max}=21$). As $\mathbf{S}$ is a vector of probabilities, the maximization must be carried out under the constraints that the $s_n$ are positive and that they add up to one. The positivity constraint can be satisfied changing variables: $s_n=\sigma_n^2$. Instead of L, we maximize the logarithm of L:

$$l(\mathbf{\Sigma}) = \ln\left(L(\Sigma)\right) = \ln(C) + \sum_{i=0}^{N} r_i \ln\left(\sum_{m=0}^{m_{max}} p_{i,m}^N \sigma_m^2\right) \qquad (5)$$

where $\mathbf{\Sigma}=[\sigma_n]$ and C is a constant.
The condition that the $s_n$ add up to one can be taken into account using the Lagrange multipliers method: $F(\Sigma,\alpha)=l(\Sigma)-\alpha\left(\sum_{m=0}^{m_{max}}\sigma_m^2-1\right)$.

After developing [35] the set of $m_{max}+2$ gradient equations $\nabla F(\Sigma,\alpha)=0$, we obtain that $\alpha=H$ and that the set of $m_{max}+1$ nonlinear equations to be solved respect to $\Sigma$ is:

$$\sigma_l \left[ \sum_{i=0}^{N} \frac{r_i p_{i,l}^N}{\sum_{m=0}^{m_{\max}} p_{i,m}^N \sigma_m^2} - H \right] = 0 \qquad (6)$$

for l=0,…, $m_{max}$. The set of equations (6) can be solved by standard numerical methods.

*7.3. ML reconstruction*

To test the effectiveness of the reconstruction algorithm, a 8.6x8 $\mu m^2$ 5-PND was tested with the coherent emission from a mode-locked Ti:sapphire laser, whose photon number probability distribution is approximately poissonian ($S(m)=\mu^m \cdot \exp(-\mu)/m!$). Therefore, **S** could be determined by measuring the mean photon number per pulse μ with a powermeter. To determine $\mathbf{Q_{ex}}$, histograms of the photoresponse voltage peak were built for varying μ. The signal from the device was sent to the 1 GHz oscilloscope, which was triggered by the synchronization generated by the laser unit. The photoresponse was sampled for a bin time of 5ps, making the effect of dark counts negligible.
The device was characterized in terms of its conditional probability matrix $\mathbf{P^5}$ ([18, 19], figure 11), so it was possible to carry out the ML estimation of the different incoming distributions with which the device was probed. Because of the bound on the number of incoming photons which is possible to represent in our algorithm ($m_{max}$=21) and as for a coherent state losses simply reduce the mean of the distribution, the ML estimation was performed considering μ*=μ/10 and **η***=10**η** (the efficiency of each section being lower than 10%)**.**

Figure 12 shows the experimental probability distribution of the number of measured photons $\mathbf{Q_{ex}}$ obtained from the histograms measured when the incoming mean photon number is μ=1.5, 2.8, 4.3 photons/pulse (figure 12 a, b, c respectively, in orange), from which the incoming photon number distribution is reconstructed. The ML estimate of the incoming probability distribution $\mathbf{S_e}$ is plotted in figure 12 d, e, f, (light blue), where it is compared to the real incoming probability distribution **S** (green). The estimation is successful only for low photon fluxes (μ=1.5, 2.8 figure 12 d, e) and it fails already for μ=4.3 (figure 12 f). In figure 12 a, b, c, $\mathbf{Q_{ex}}$ (orange) is compared to the ones obtained from **S** and $\mathbf{S_e}$ through relation (2) (**Q**, $\mathbf{Q_e}$ in green and light blue, respectively).

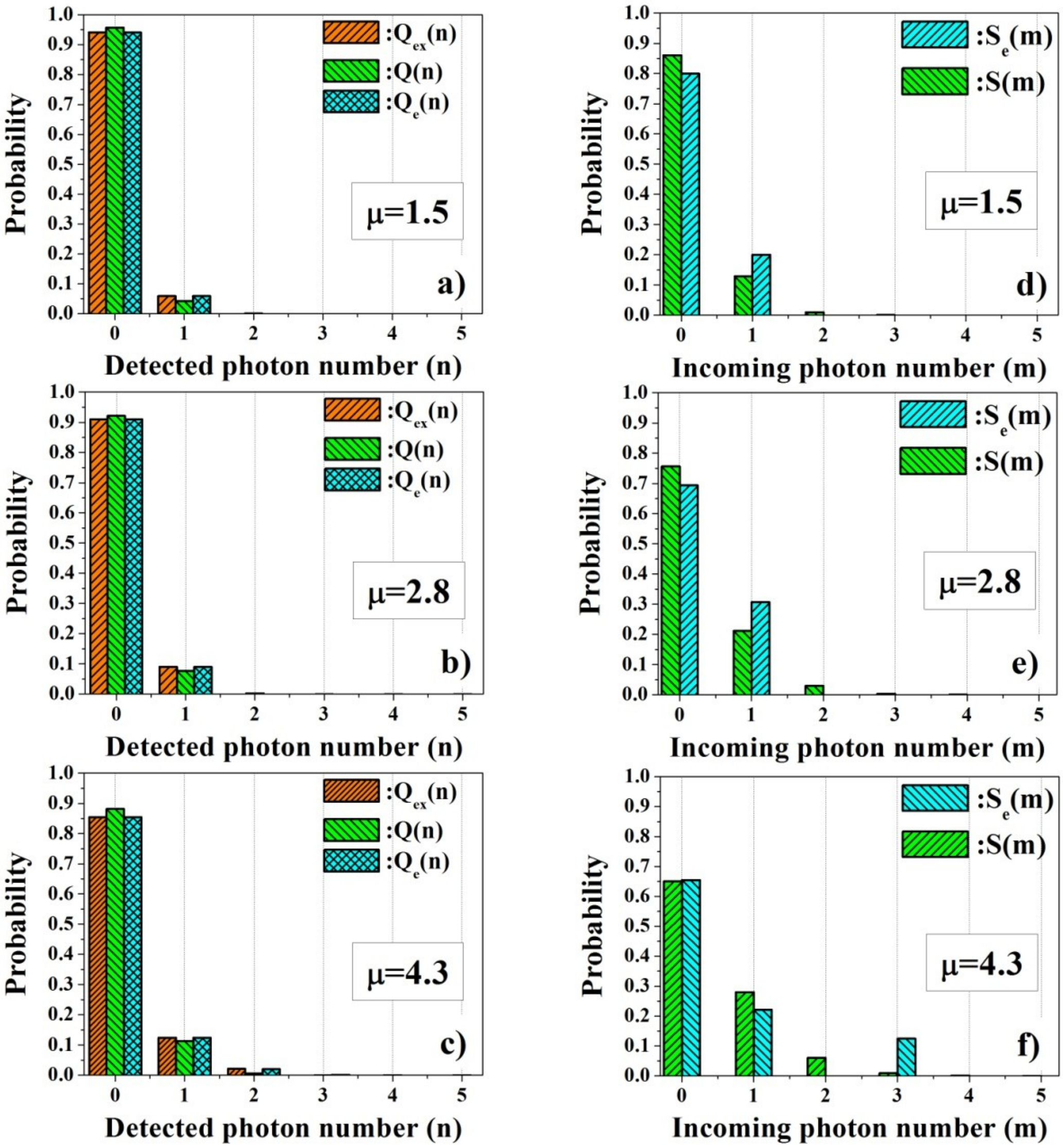

**Figure 12. (a), (b), (c) Probability distribution of the number of measured photons obtained from experimental data $Q_{ex}$ (orange), from S (Q, in green) and $S_e$ ($Q_e$, in light blue) through relation (2) for μ=1.5, 2.8, 4.3 photons/pulse, respectively. (d), (e), (f) Real incoming probability distribution S (green) and its ML estimate Se (light blue) for μ=1.5, 2.8, 4.3 photons/pulse, respectively. The 8.6x8 µm2 5-PND was tested under uniform illumination in a cryogenic dipstick dipped in a liquid He bath at 4.2 K. The light pulses at 700nm form a mode-locked Ti:sapphire laser were 40ps wide (after the propagation in the optical fiber) and the repetition rate was 80MHz. The average input photon number per pulse µ was set with a free space variable optical attenuator.**

The main reasons why the reconstruction fails are not the low efficiencies of the sections of the PND or the the spread in their values, but rather the limited counting capability (N=5) and a poor calibration of the detector, i.e. an imperfect knowledge of its real matrix of conditional probabilities. This assessment is supported by the following argument. If we generate $\mathbf{Q_{ex}}$ with a Montecarlo simulation [19] using the same $\boldsymbol{\eta}$ vector of Figure 11 and a poissonian or thermal incoming photon number distributions and then we run the ML reconstruction algorithm (using the same $\mathbf{P^5}$, which this time describes perfectly the detector), **S** can be estimated up to much higher mean photon numbers. However, to alleviate this problem, a self-referencing measurement technique might be used [36].

## 8. Conclusion

A new PNR detector, the Parallel Nanowire Detector, has recently proven to significantly outperform existing approaches in terms of sensitivity (NEP=$4.2\times10^{-18}$ W/Hz$^{1/2}$), speed (80 MHz count rate) and multiplication noise [18, 19] in the telecommunication wavelength range.

An electrical equivalent circuit of the device was developed in order to study its operation and to perform its design. In particular, we found that the leakage current significantly affects only the PND detection efficiency, while it has a marginal effect on its signal to noise ratio. Furthermore, in order to gain a better insight on the device dynamics, the (N+1)-mesh equivalent circuit of the N-PND was simplified and reduced to a three mesh circuit, so that the analytical expression of its transient response could be easily found. With this approach, we could predict a physical limit to the recovery time of the PND, which is slower than that previously estimated. Additionally, the figures of merit of the device performance in terms of efficiency, speed and sensitivity ( $\overline{\delta I_{lk}}$ , $f_0$, SNR) were defined and their dependency on the design parameters ($L_0$, $R_0$, $R_{out}$, N) was analyzed.

In order to prove the suitability of the PND to reconstruct an unknown light statistics by ensemble measurements, we developed a maximum likelihood estimation algorithm. Testing a 5-PND with a Poissonian light we found that the reconstruction of the incoming photon number probability distribution to be successful only for low photon fluxes, most likely due to the limited counting capability (N=5) and the poor calibration (i.e. the imperfect knowledge of the real matrix of conditional probabilities) of the detector used, and not to its low detection efficiency ( $\tilde{\eta} \approx 3\%$ ). Additional simulations will be needed to evaluate the performance of our detector for the measurement of other, nonclassical photon number distributions. Finally, despite the high sensitivity and speed of PNDs, their present performance in terms of detection efficiency ($\eta$=2% at $\lambda$=1.3μm) does not allow their application to single shot measurements, as required for linear-optics quantum computing [4], quantum repeaters [5] and conditional-state preparation [6]. Nevertheless, the $\eta$ of SPDs based on the same detection mechanism can be increased to ~60% [29], and could potentially exceed 90% using optimized optical cavities.

## 9. Acknowledgments

This work was supported by the Swiss National Foundation through the "Professeur boursier" and NCCR Quantum Photonics programs, EU FP6 STREP "SINPHONIA" (contract number NMP4-CT-2005-16433), EU FP6 IP "QAP" (contract number 15848). The authors thank B. Deveaud-Plédran and G. Gol'tsman for the useful discussions, B. Dwir and H. Jotterand for technical support and the Interdisciplinary Centre for Electron Microscopy (CIME) for supplying SEM facilities. A. Gaggero gratefully acknowledges a PhD fellowship at University of Roma TRE.